# Spinal Coupling in Frontal and Transversal Plane During Gait – A Segmental and Time-Dependent Analysis of the Thoracic and Lumbar Spine

Jonas Dully[1*], Carlo Dindorf[1], Henriette Rönsch[2], Dennis Perchthaler[1], Jürgen Konradi[2], Ulrich Betz[2], Michael Fröhlich[1]

[1]Department of Sports Science, RPTU University Kaiserslautern-Landau, Kaiserslautern, Germany

[2]Institute of Physical Therapy, Prevention and Rehabilitation, University Medical Center of the Johannes Gutenberg University Mainz, Mainz, Germany

*Corresponding author: Jonas Dully (jonas.dully@rptu.de)

## Abstract

**Introduction:** Coupling between lateral deviation and axial rotation is a known feature of spinal mechanics, yet its behavior at the individual vertebral level during gait, as well as the association with sagittal posture remains poorly understood. **Methods:** This study analyzed spinal kinematics in a diverse cohort (n=642) using a non-invasive rasterstereography system with an instrumented treadmill, quantifying time-dependent coupling of rotation and lateral deviation for each vertebra from T3 to L4 during walking as well as the influence of sagittal posture on this coupling. Coupling behavior was analyzed with absolute phase lag, normalized signed area and tilt angle derived from the Fourier series. **Results:** Results revealed cranio-caudal patterns for all three metrics with differences between almost all adjacent vertebrae and different turning points, i.e., where the coupling behavior changed (from increase to decrease and vice versa). Generalized, as well as pooled static sagittal posture significantly modulated these patterns, while individual deviations from static sagittal posture influenced the metrics. **Conclusion:** These findings provide the first dynamic, vertebra-level characterization of spinal coupling during gait. Furthermore, they offer a foundation for an understanding of spinal biomechanics and the influence of static sagittal posture, potentially relevant to the diagnosis and treatment of spinal disorders.

## Introduction

Spinal movement is highly individual [1] and dependent on pathologies and restrictions [2,3], as well as overall anthropometric characteristics [4,5]. With regard to the high prevalence of back pain in Western societies [6], its impact on quality of life and disability [7], and its socioeconomic burden [8], a deeper understanding of spinal biomechanics is of considerable clinical relevance. Despite extensive research, spinal kinematics remain incompletely understood because of the complex three-dimensional interactions and coupling mechanisms between vertebral bodies [9], even during everyday activities such as walking [10].

Previous cadaveric and in vivo studies have demonstrated coupling between vertebral motions both within individual anatomical planes (between different vertebrae) and across different planes, particularly between lateral flexion and axial rotation (within a vertebra) [9,11,12]. However, these may not represent the realistic in vivo movements, since cadaver studies neglect muscle and tendon dynamics [13], while other studies are limited by the use of highly standardized rotational tasks that do not reflect activities of daily living [11,14,15].

Coupled motion between frontal-plane and transverse-plane movements has been consistently reported during standardized spinal movements [9,11,12]. Nevertheless, the direction and magnitude of this coupling appear to depend on spinal region and body posture [14,15]. While a strong association between axial rotation and ipsilateral lateral flexion has been described for the subaxial cervical spine [15], findings for the thoracic spine remain inconsistent. Previous studies have reported ipsilateral coupling patterns [16,17], variable or segment-dependent behavior [18,19], as well as contralateral flexion during rotational movements [20]. These contradictory findings highlight the need for further investigations of thoracic spine kinematics under physiologically relevant in-vivo conditions.

Also, there is evidence that the sagittal posture of the spine does affect the coupling in frontal and transversal plane [15]. This may be attributed to differences in vertebral morphology and orientation, particularly of the spinous processes [21], which influence facet joint kinematics by altering the available degrees of freedom within the neutral zone, i.e., the area in which the vertebra can move with minimal internal resistance [22].

To the authors' best knowledge, only one study investigated the coupling between lateral bending and rotation of the spine during gait [12], while no study investigated the coupling of lateral bending and rotation of each vertebral body during gait. Furthermore, the influence of sagittal posture on the coupling has only been reported on spinal segments and provoked bending forwards or backwards [15].

Based on these research gaps, this study addresses two main research questions. First (research question (RQ1)), how does the time-dependent coupling between frontal and transversal plane motions change along the cranio-caudal axis. Second (RQ2), to what extent does sagittal orientation moderate this coupling across individual vertebrae?

By quantifying vertebra-specific coupling patterns and their relationship to sagittal vertebral orientation during walking, this study advances current understanding of three-dimensional spinal kinematics and provides novel insights into the mechanisms underlying regional spinal coordination.

## Methods

*Overview and theoretical background*

Following the visualizations of [12] and [23], coupled spinal motion was characterized using angle-angle plots subsequently named hysteresis curves, due to their characteristic shapes and time-dependency), derived from the relationship between frontal-plane and transverse-plane vertebral motion during gait.

To quantitatively characterize these coupling behaviors, three complementary metrics were evaluated. Temporal offset between lateral deviation and axial rotation was quantified using the *absolute phase lag* (|ϕ|) (APL). Coupling magnitude, divergence and temporal precedence of motion initiation were assessed using the *signed normalized loop area (SNA)*. Directional relationship, reflecting ipsilateral or contralateral relationships between lateral deviation and axial rotation, was quantified using the *Fourier-derived coupling tilt angle (θ) (FTA)*. For a detailed interpretation of these metrics and their biomechanical implications, please refer to Table 1.

*Participants, data and preprocessing*

Gait and static standing data were obtained using rasterstereography, including plantar pressure measurements with an instrumented treadmill (DIERS 4D motion® Lab, Wiesbaden, Germany), and analyzed using the corresponding software (DICAM v23.03.0.18). Data were restricted to participants with complete measurements across all vertebral levels and collected under naturalistic clinical conditions. The data were provided by the manufacturer in an anonymized format and comprised spinal kinematics as well as year of birth, year of data acquisition, and biological sex as the only personal information, with the manufacturer ensuring compliance with the applicable requirements for data transfer. This retrospective analysis was conducted in accordance with the ethical approval granted by the responsible ethics committee (Approval No. 106). As only the birth year and year of data acquisition were available, age was defined as the participant's age at the end of the acquisition year. Consequently, only participants aged at least 19 years were included to ensure all participants were legal adults. This resulted in a population of 642 participants (biological sex: 375 w, 267 m, age at the end of the year of acquisition: 45.3 ± 12.9 years; gait velocity: 3.40 ± 0.53 km/h).

From the rasterstereography measurements, spinal kinematics in the frontal and transverse planes from each vertebral body (vertebra prominens (VP)-L4). Lateral deviation (frontal plane kinematics of the vertebrae) was defined as the lateral displacement of the vertebral foramen from the spinal midline, extending from the midpoint between the posterior superior iliac spine dimples to the vertebra prominens (C7). The parameter was automatically calculated by the manufacturer's software according to the system-specific algorithm and has been described previously [11]. Rotation is defined as the rotational movement in the transversal plane around the foramen of each vertebra in contrast to the pelvis. Sagittal deviation is defined as the deviation in sagittal plane of the vertebra to the vertical line between the mid of the dimples to VP.

For the plantar pressure data, registration of the foot side was applied, and the resulting ground reaction force was filtered using a $4^{th}$ order Butterworth low pass filter with a cut-off frequency of 12 Hz, while the kinematical data were low-pass filtered using a zero-lag 4th-order Butterworth filter with a cut-off frequency of 6 Hz [24]. Filtering was performed using a bidirectional implementation in second-order sections to ensure numerical stability and eliminate phase shift [25]. Heel strike and toe-off were segmented, with the vertical ground reaction force, when the resulting force was above 5 N and each gait cycle was time-normalized to 101 time-points (0-100%). The segmented gait cycles were averaged per participant to achieve a representative gait pattern of each participant, consistent with previous analyses using comparable data [26].

Lateral deviation of each vertebra was standardized to the width of the dimples from the static trial to account for different body dimensions anthropometrics. Afterwards, each waveform was centered by subtracting its mean value to isolate the dynamic coupling behavior independently of the absolute vertebral position or subject-specific pre-rotation (e.g., caused by scoliosis). This approach was chosen because the average spinal kinematic profile closely reflects an individual spinal profile [27], whereas the present study focuses on identifying generalized coupling characteristics at the population level.

*Coupling metrics*

**Table 1.** Description, interpretation and calculation of the metrics, including their possible ranges (ROT refers to rotation and LD refers to lateral deviation). DOF = direction of freedom; LD = lateral deviation; ROT = rotation

| Hysteresis characteristics | | | | |
|---|---|---|---|---|
| Variable | Description | Interpretation | Formula | Range |
| Absolute Phase lag (ϕ) [°] | Temporal offset between LD(t) and ROT(t) at the first harmonic.<br>Quantifies how far apart in phase the two rotational DOFs are, regardless of which leads. | Larger \|ϕ\|: greater temporal separation of the DOFs<br>Near 0°: the DOFs move nearly synchronously.<br>Near 90°: one DOF leads the other, while the other starts when the first does not anymore - maximal phase separation.<br>Near 180°: the two DOFs move nearly synchronously. | $\|\varphi\| = \left\|\arctan\left(-\frac{s^1}{c^1}\right)\right\|$<br>where $s_1$, $c_1$ = sine and cosine coefficients of the first Fourier harmonic of LD(t) | [0° – 180°] (unsigned) |
| Signed normalized area (A) | Loop fill relative to its bounding box. 0 = collapsed to a line (perfectly in-phase). ~0.785 = perfect ellipse. 1 = filled rectangle. | A > 0: Loop is traversed counter-clockwise: lateral deviation leads rotation. In the very first moment, the spine bends and then rotates.<br>A < 0: Loop is traversed clockwise: rotation leads lateral deviation. In the very first moment, the spine rotates and then bends.<br>A = 0: The curve is collapsed to a line | $A = \frac{\|A_{shoelace}\|}{w * h} * sign(A_{shoelace})$<br>where:<br>$A_{shoelace} = \frac{1}{2}\sum_i (x_i y_{i+1} - x_{i+1} y_i)$<br>- the shoelace formula applied to the closed loop<br>$w = \max(x) - \min(x)$<br>- lateral deviation range [mm]<br>$h = \max(y) - \min(y)$<br>- rotation range [°] | [-1, 1] |
| Fourier tilt angle [°] | Orientation of the first-harmonic ellipse in the lateral deviation–rotation plane, derived analytically from the Fourier amplitudes (as described above) and phase of the two coupled signals. | ψ > 0: axis tilts into first and third quadrants of the ROT×LD plane. Right ROT, leads to right LD (left ROT leads to left LD) — ipsilateral coupling.<br>ψ < 0: axis tilts into second and fourth quadrants. Right ROT accompanies left LD (left ROT leads to right LD) — contralateral coupling.<br>ψ ≈ 0: loop's long axis runs along ROT axis with minimal LD component. Coupling is nearly pure ROT with little LD.<br>Ψ ≈ {-90, 90}: loop's long axis runs along LD axis with minimal ROT. Coupling is nearly pure LD with little ROT. | $\psi = ½ \arctan\left(\frac{2A_{ROT}A_{LD}\cos\Delta\varphi}{(A_{ROT}^2 - A_{LD}^2)}\right)$<br>where $A_{ROT}$, $A_{LD}$ = first-harmonic amplitudes and $\Delta\varphi = \varphi_{ROT} - \varphi_{LD}$ = phase difference, identical to those used for APL computation. | [−90°, +90°] (from arctan) |

Hysteresis magnitude and shape were summarized with three operational metrics: The first-harmonic phase lag between lateral deviation and rotation (APL), a normalized hysteresis-loop area measure capturing energy dissipation and loop shape (SNA), and the Fourier-derived tilt angle of the fitted first-harmonic ellipse in the lateral deviation–rotation plane (FTA), which were derived from methods and ideas in analysis of angle-angle plots in biomechanics or hysteresis analysis in general [28–31]. This was done for each subject.

For visualization, the bootstrapped mean and 95% confidence intervals were calculated for each vertebra and timestep (percent) with a bootstrap seed of 1000. All of the above and below described analysis (except for statistics) were done in Python (version 3.13) using NumPy [32], while visualizations were done with matplotlib [33] and seaborn [34].

*Statistical analysis*

For each subject, hysteresis loop metrics (APL, SNA, and FTA) and static sagittal deviation were determined at 14 vertebral levels (T3–L4), leading to one measurement per metric per vertebra, corresponding to a repeated-measures design with vertebra nested within subject and the sagittal deviation as a between factor. To avoid collinearity between the vertebra factor and sagittal deviation, the latter was centered with a group-mean (vertebra)- centering. For each vertebral level, the subject's sagittal deviation was centered on that level's own sample mean.

Following, for each metric, a linear mixed-effects model was fitted with vertebra (14-level categorical factor), within-subject sagittal deviation, and their interaction as fixed effects, and a random intercept for subject.

$$metric = vertebra\ x\ flexion + (1|subject_{id})$$

Models were fitted by restricted maximum likelihood (REML) using lme4 [35] *and* lmerTest [36] in R (version 4.5.3) [37], and evaluated with type-III F-tests using the Kenward-Roger approximation for denominator degrees of freedom with the pbkrtest package [38].

The main effect vertebra and the interaction effects were post-hoc-analysed using estimated marginal means using the emmeans package [39]. Pairwise comparisons between anatomically adjacent vertebrae were evaluated using a pairwise t-test on the estimated marginal means, with Bonferroni correction for the 13 adjacent comparisons ($\alpha = .05$).

For each vertebral level, the linear mixed-effects model described above was used to test whether an individual's own sagittal deviation was associated with the metric. i.e., whether subjects whose posture at that level were more or less deviated than typical showed correspondingly different metric values. The metric was predicted at the 10$^{th}$ and 90$^{th}$ percentile of within-subject sagittal deviation observed at that level, and the difference between these two predictions (high minus low) was tested using estimated marginal means. The resulting p-values were Bonferroni-corrected ($\alpha = .05$). Predicted postures were converted back to raw sagittal deviation by adding each level's own mean. Standardized effect sizes (Cohen's *d*) were computed for all pairwise and moderation contrasts as the raw estimate divided by the model's residual standard deviation.

To assess whether each metric's spatial pattern across vertebral levels corresponded to the population-level sagittal deviation profile, the per-vertebra mean metric value was correlated with the per-vertebra mean sagittal deviation (Pearson and Spearman, n=14 levels). For comparison, the same correlation was additionally computed across all subject-level observations pooled across vertebrae. Because this pools 14 non-independent measurements per subject, the resulting p-value does not satisfy the independence assumption underlying the correlation test; this pooled value is reported exploratively only and was not used for inferential claims.

Exploratory robustness analyses demonstrated that the generalized coupling patterns adequately represented individual trajectories. In addition, APL values obtained from first-order ($K = 1$) and third-order ($K = 3$). Fourier approximations were nearly identical, supporting the use of the simpler first-order model for subsequent analyses.

## Results

### *Generalized visualization, characterization of coupling*

Figure 1a and 1b. illustrate the hysteresis curves including the metrics for each vertebra respectively. Visually there are differences between the coupling of the lateral deviation and rotation for most of the vertebra. These differences were subsequently quantified and statistically analyzed using the introduced hysteresis curve–based metrics and statistical tests.

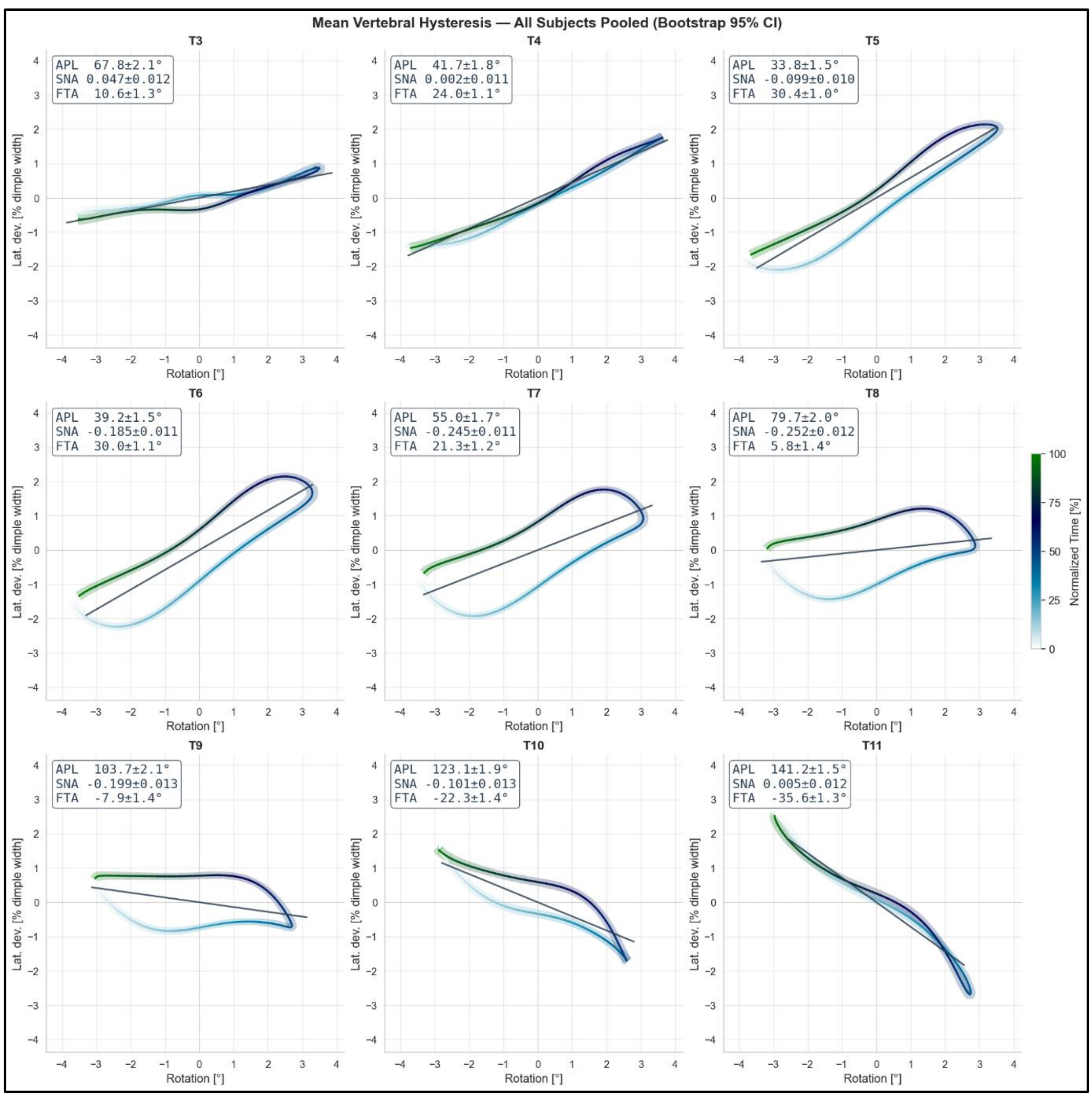

**Figure 1a.** Coupling curves (bootstrapped mean ± 95% confidence interval) for T3-T11, including median metrics ± quartiles. The crossing lines represent the angle derived from the tilt of the first Fourier harmonic (FTA)

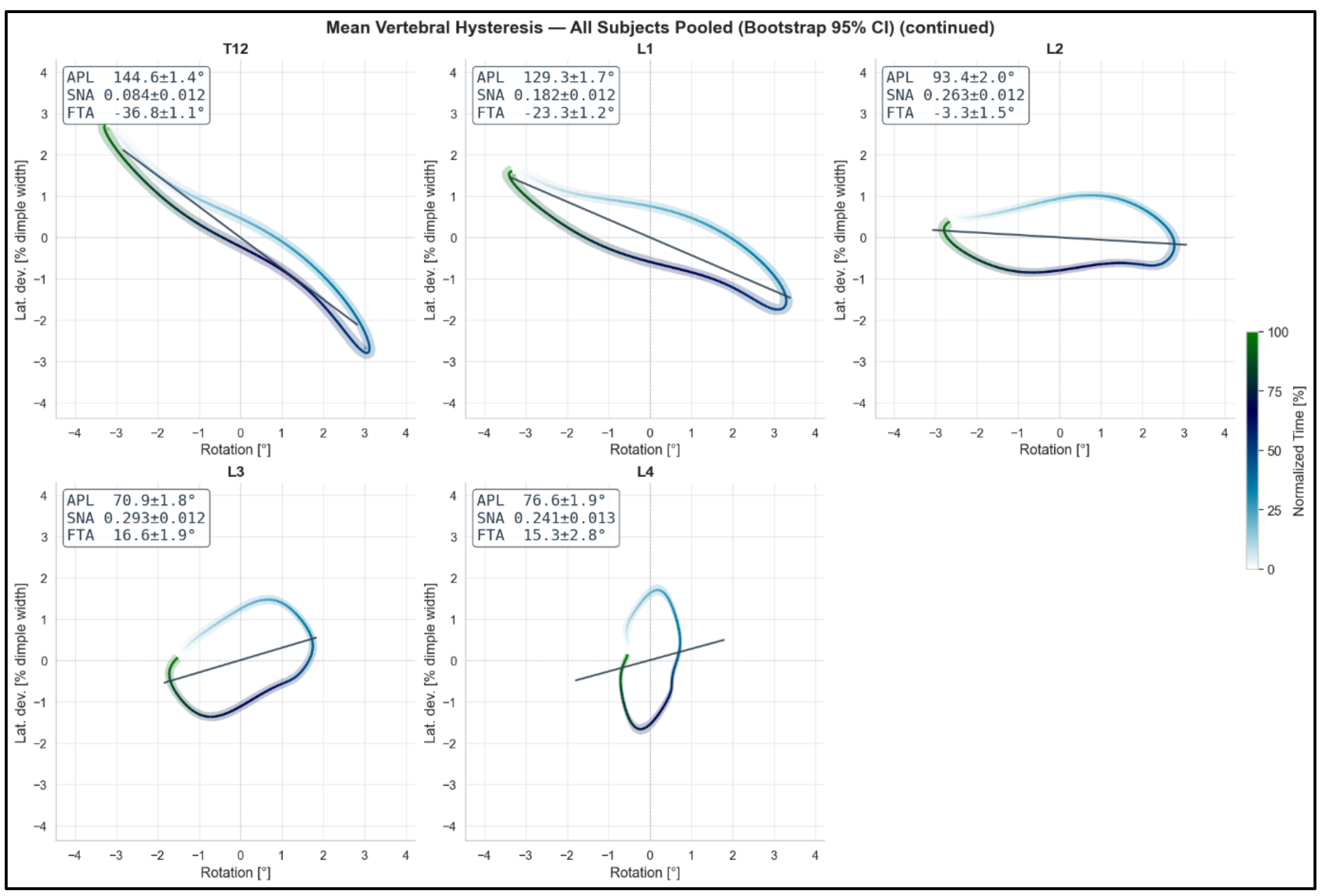

**Figure 1b**. Coupling curves (bootstrapped mean ± 95% confidence interval) for T12-L4, including median metrics ± quartiles. The crossing lines represent the angle derived from the tilt of the first Fourier harmonic (FTA)

Shown in Figure 2. all main and interaction effects were significant, while the shape correlation is significant with large effect for APL and FTA, and not statistically existent for SNA.

Across the cranio-caudal axis, all metrics exhibited pronounced vertebra-specific changes in coupling characteristics. APL showed a biphasic pattern, decreasing from T3 to a trough at T5, increasing to a peak at T12, and subsequently declining to a second trough at L3 before increasing again toward L4. SNA decreased from T3 to a trough at T8, followed by a gradual increase to L3 and a subsequent decline toward L4. In contrast, FTA increased to a peak at T5, declined to a trough at T12, increased again toward L3, and subsequently decreased toward L4.

*Hysteresis characteristics*

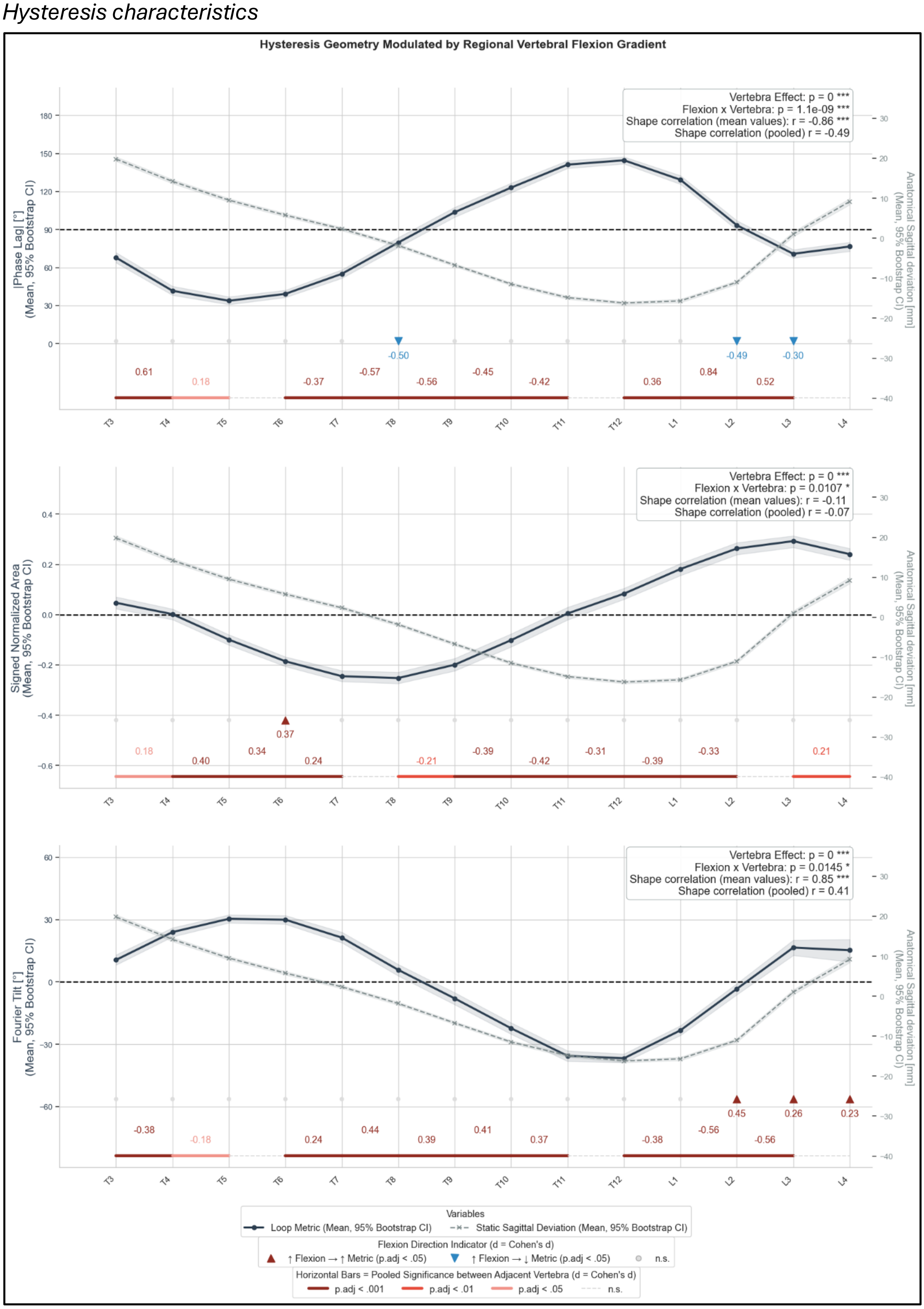

**Figure 2.** Distribution (mean ± bootstrapped CIs) of the hysteresis coupling metrics (absolute phase lag, signed normalized area, and Fourier tilt) across vertebral levels, together with the corresponding post hoc analyses. Horizontal lines indicate differences between adjacent vertebral levels, whereas triangles

indicate differences and direction of differences in individual sagittal deviation within vertebral levels. Effect sizes for differences (Cohen's d) between adjacent vertebral levels are shown above and below the horizontal lines, and effect sizes for profile differences within vertebral levels are shown above the triangles.

Inferential statistics show that for APL, all adjacent vertebrae showed to be different, but between T5 to T6, T11 to T12 and L3 to L4 with absolute very small (*d* = 0.18 ) to large (*d* = 0.84) effect sizes. For the SNA, only T7 to T8 and L2 to L3 were not significant, with the other effects showing very small effect sizes (*d* = 0.18 (very small, T3-T4) while other absolute effect sizes while others were moderate (*d* = [0.2, 0.42]). For FTA, T5 to T6, T11 to T12 and L3 to L4 were not significant, with the others also showing very small (*d* = 0.18 at T4 to T6) to moderate absolute effect size (maximal *d* = 0.56 at L to L2). All these effect sizes are interpreted according to Cohen [40].

Regarding the moderation of the sagittal deviation for the effect within the vertebra, some interactions were found in the post-hoc-tests. For APL, at T8, L2 and L3, moderate negative effects were observed (*d* = [-0.30, -0.50]), showing that with increasing sagittal deviation, the metric decreases. For SNA, only one effect was found for T6 with moderate positive effect size (*d* = 0.37), which tells that with increasing sagittal deviation, SNA increases at T6. Regarding FTA, moderate positive effect sizes were found at L2 (*d* = 0.45), L3 (*d* = 0.26) and L4 (*d* = 0.23).

## Discussion

The present study presented the coupling behavior in the transversal and frontal plane of the spine on a vertebra-level (T3-L4) derived from rasterstereographic data during gait. Additionally, this study evaluated the effect of static sagittal posture on this coupling. Several interesting findings can be described.

First and foremost, the frontal-transversal coupling differs both between most vertebral levels, indicating that spinal coordination is not uniform along the spine (RQ1). Previous studies have already reported vertebra-dependent differences in spinal rotation during gait [41], although these differences were mainly observed in the lumbar spine, while the thoracic spine primarily differed in amplitude rather than waveform timing or shape. Differences along the cranio-caudal axis have been reported before, but either did not consider gait or functional movements at all or were performed on spine cadavers [11,13–15]. Therefore, this study is the first to characterize this coupling behavior on a vertebra-level during gait.

Across all coupling metrics, pronounced cranio-caudal changes in spinal coupling behavior were observed, supporting the presence of region-specific coordination patterns during gait. For APL, there was a progressive decrease from the upper thoracic to the mid-thoracic spine and an increase toward the thoracolumbar junction, followed by a decrease toward the lumbar spine. Significant differences between all adjacent vertebrae, except for T5-T6, T11-T12 and L3 to L4 were found. These findings indicate that the temporal synchrony between lateral deviation and axial rotation changes along the cranio-caudal axis, with maximum asynchrony in the T8/T9 and L2 regions and increasing synchrony toward T5 and T12.

Similarly, SNA demonstrated substantial cranio-caudal changes in coupling behavior. Significant differences between adjacent vertebral levels were observed throughout most of the spine (except for T7-T8 and L2-L3), indicating continuous changes in which direction leads the coupled movement. In regions T4 and T11, the differences are minimal, while the lead of rotation reaches its maximum toward T8 and the lead of lateral deviation reaches its maximum toward L3. Further, in the lumbar and mid-thoracic (T6-T8) spine, the normalized area peaked, where the angle-angle plots show greater areas, while in T3-T5, T11 and T12, the plots are almost degenerated lines closing around almost no area.

For FTA, nearly all adjacent vertebral levels differed significantly, demonstrating continuous cranio-caudal changes in coupling directionality (except for T5-T6, T11-T12 and L3-L4). Ipsilateral coupling between lateral deviation and axial rotation was observed from T3–T8, which is consistent with previous literature describing ipsilateral thoracic coupling behavior [16,17]. In contrast, T9–T12 demonstrated contralateral coupling, which corresponds to previous reports describing variable or contralateral thoracic movement patterns [18–20,42].

For L1 and L2, the present findings agree with previous literature describing contralateral lateral deviation during axial rotation [43]. However, unlike previous studies reporting contralateral coupling at L3 [11,44,45], the present study demonstrated ipsilateral coupling at this vertebral level. One possible explanation is that previous studies primarily investigated standardized upper- or lower-body movement tasks, whereas the present work focused on gait. During walking, pelvic motion occurs simultaneously in multiple degrees of freedom and contributes to both spinal rotation and lateral deviation [24,46], resulting in a complex interaction between these movement planes where direct causal relationships cannot easily be isolated. In addition, gait introduces coordinated counter-rotation between the pelvis and shoulders [47,48], a dynamic interaction that is not considered in most studies investigating isolated trunk movements.

In particular, the hysteresis curves show that the distinction between ipsilateral and contralateral couplings is merely a simplified categorization of the actual complexity and time-dependency of the movements. In fact, it is apparent that each individual vertebral body is integrated in a highly individualized manner into the extremely complex whole-body gait movement. The function of adjacent segments changes — much like the anatomy of the vertebral segments — gradually and without abrupt changes.

This study also evaluated the influence of static sagittal posture on the coupling in the frontal and transversal plane (RQ2). First and foremost, APL (correlation of mean curves:

r = -0.86; pooled correlation: r = -0.49) and FTA (correlation of mean curves: r = -0.85; pooled correlation: r = -0.41) moderately to strongly correlated with the static sagittal posture. A previous study also described that the coupling behavior is dependent on the overall static posture [15]. This phenomenon again suggests that the function of the spine is strongly influenced by its anatomy, which harmoniously adapts the size and shape of its structural elements along the craniocaudal axis and along the curves in the sagittal plane. This is evident from the fact that the superior articular processes of T11 and T12 are positioned like those in the thoracic spine, whereas the inferior articular processes are positioned like those in the lumbar spine [49]. Under static conditions in neutral lumbar position, rotation and lateral flexion are coupled contrary [49]. In our analysis the direction of coupling (FTA) changed in the lumbar spine. While T9 till L2 showed a contralateral coupling, L3 and L4 demonstrated ipsilateral coupling. One reason could be the anatomical conditions of L3: In contrast to L4 and L5, the vertebral motion of L3 is not restricted due strong iliolumbar ligaments and additionally mobile due insertions of dorsal latissimus and spinalis muscles [49]. Furthermore, a previous study has shown that the change in the direction of rotation does not occur at fixed vertebral body planes, but rather as a time-delayed process that begins in the lower lumbar spine/pelvis and continues cranially during gait [41].

In our point of view, the APL roughly described an inversed sagittal profile and the FTA roughly a sagittal posture along the cranio-caudal axis, which means that the more pronounced the coupling, the greater the synchrony and /or vice versa. To the best of the authors' knowledge, no data on this topic has been reported in the scientific literature to date.

This study also dived deeper and evaluated how individual deviations from the sagittal posture influences the coupling behavior. The APL shows lower values with higher sagittal deviation at T8, L2 and L3. In consequence, persons with higher thoracic kyphosis (T8) and less lumbar lordosis (L2, L3) showed less temporal asynchrony in lateral deviation and rotation. To the authors' best knowledge only one study investigated the effect of sagittal posture on the coupling [15]. However, in this work, only the overall spine was in focus, and the posture was intentionally differed, which departs from this present work, where the individual, habitual sagittal posture was considered. Therefore, the current results are standalone and can only be seen as such. In depth, the results describe that the temporal offset between lateral deviation and rotation decreases (more synchronous) when the sagittal deviation increases at T8, L2 and L3 (APL). Furthermore, the amplitude and offset between the planes in scope at T6 increases with increasing sagittal deviation (SNA). Last, the extent of the coupling direction at L2, L3 and L4 increases in the direction of ipsilateral coupling with greater sagittal deviation (FTA).

By integrating the course of vertebra-specific coupling along the cranio-caudal axis (RQ1) with the effect of overall and individual sagittal posture (RQ2), distinct *transition regions* (where the sign of the gradient changes) in frontal–transversal spinal coupling were observed. Across all coupling metrics, qualitative changes (local extrema) in coupling behavior were consistently observed in the mid-thoracic region (approximately T5 and T8) and at the thoracolumbar junction (T12), as well as the lumbar spine (L3). These findings suggest that spinal coordination during gait is regionally organized and oriented on the anatomy, rather than mechanically uniform.

This regional organization is comparable with previous gait studies reporting maximal rotational amplitudes around T7–T8 and progressive shifts in rotational timing toward the lower thoracic and lumbar spine [41]. Functional analyses have additionally described T7 as a transitional region with minimal frontal-plane motion relative to the pelvis during gait [41,50]. The present results extend these observations by demonstrating that these transition zones are also reflected in the dynamic coupling between lateral deviation and axial rotation.

The goodness of the fit between the mathematical description and the data (as well as the median fit for all subjects), quantified by $R_{xy}^2$ (0.28-0.56), and $R_{xyS}^2$ (0.40-0.65) shows that the coupling is mathematically quite well generalizable. Although a few vertebral levels in the lumbar spine showed weaker fits to their mathematical description, most demonstrated moderate agreement between the mathematical model and both pooled and participant-specific trajectories. Notably, the similarity between overall and subject-level fits suggests that the population-level representation adequately captures individual coupling behavior. Considering the known individuality of spinal motion during gait [1], these findings support the existence of consistent underlying coupling patterns [9] despite substantial interindividual variability.

Several limitations should be considered when interpreting the present findings. First, vertebral body motion was estimated indirectly from fine-grained rasterstereographic surface topography rather than measured directly. Nevertheless, the system has demonstrated validity and reliability in spinal parameters during standing [51–54], and good agreement in dynamic reconstruction of the back surface in standardized movements with gold-standard biomechanical and radiographic methods, while remaining non-radiating [55]. In gait analysis, this system is reliable [1,26], however, no validation was performed during walking tasks. Second, data were collected under naturalistic clinical conditions, resulting in a heterogeneous patient sample that may have influenced specific effects, particularly as spinal positioning and deviation is known to affect coupling behavior [15,56]. At the same time, the persistence of the observed coupling patterns across a heterogeneous population may support the robustness and generalizability of the findings. Furthermore, age-related differences in spinal and trunk kinematics reported in previous studies [57–59] were not analyzed due to limited subgroup sizes and should be addressed in future research.

## Conclusion

This study presents the first vertebra-level analysis of lateral deviation–axial rotation coupling during gait, revealing a systematic cranio-caudal organization with distinct ipsilateral and contralateral coupling zones and several transition points along the spine. Importantly, these results are moderated by the sagittal posture of the overall spine and in some extent by the individual sagittal deviation of some vertebrae.

These findings extend current understanding of spinal coupling beyond standardized tasks, cadaveric studies and fixated postures and suggest that spinal coordination during walking is regionally specialized rather than mechanically uniform. The results provide a biomechanical foundation for future research into spinal disorders and therapeutic approaches targeting vertebra-specific spinal mechanics with consideration of the sagittal profile.

## Author contributions

Conceptualization: JD, CD, HR, DP, JK; Data curation: JD, CD, HR; Formal analysis: JD, CD; Funding acquisition: CD, JK, DP, UB, MF; Investigation: HR, JK; Methodology: JD, CD, MF; Project administration: CD, DP, UB, MF; Resources: DP, JK, UB, MF; Software: JD, CD, DP, MF; Supervision: UB, MF; Validation: JD, CD, HR, JK; Visualization: JD; Writing – original draft: JD, CD, HR, DP, JK; Writing – review & editing: UB, MF

## Data availability statement

Data will be made available upon reasonable request

## Competing Interests Statement

None of the authors has any competing interests

## Funding

This research was supported by the Central Innovation Program for Small and Medium-Sized Enterprises (Zentrales Innovations Program Mittelstand, ZIM) of the German Federal Ministry for Economic Affairs and Climate Action under Grant numbers 16KN113027, 16KN113026 and KK5209402NK4.